\documentclass[11pt,letterpaper]{JHEP3}
\usepackage{graphicx}
\usepackage{amsmath}
\usepackage{epsfig}

\title{A vector-like heavy quark in the Littlest Higgs model}
\author{Jaeyong Lee\\ 
Department of Physics, Box 1560, 
University of Washington, Seattle, WA 98195-1560 \\
E-mail: \email{jaeyong@u.washington.edu}}
\preprint{UW/PT-04-13}
\abstract{The Littlest Higgs model contains a new vector-like heavy quark 
in the up sector. There are two interesting features of its 
existence. One is that it extends the $3\times3$ CKM matrix 
in the Standard Model to a $4\times3$ matrix and
the other is that it allows Z-mediated flavor changing neutral currents 
at tree level in the up sector but not in the down sector. 
We examine a few of possible windows in which the Z-mediated flavor 
changing neutral currents in the Littlest Higgs Model can be tested.}
\keywords{hig}

\begin{document}

\section{Introduction}
A variety of experiments has revealed three generations for quarks
in the Standard Model. Theoretically the cancellation of the gauge anomalies
explains the equal number of weak doublets of quarks and leptons. 
This also account for three quark generations.
The number of quark generations implies the $3\times3$ 
Cabibbo-Kobayashi-Maskawa (CKM) matrix for the quark mixing.
Weak decays of the relevant quarks or deep inelastic neutrino scattering
can determine all the values of the CKM  matrix elements.
Therefore, the number of quark generations, in principle,
can be tested by measuring the CKM matrix.
So far the experimental data have not precluded 
there being more than three generations.

A new theory for electroweak symmetry breaking was developed
from deconstruction theory two years ago\cite{Arkani-Hamed:2002qy,Han:2003wu}.
This theory is named as ``the Littlest Higgs (LH) model''
because it is the smallest extension of the SM to date that stabilizes 
the mass of the SM Higgs boson on the electroweak scale.  
The SM Higgs doublet belongs to a Goldstone multiplet in an $SU(5)/SO(5)$ 
nonlinear sigma model. Other elements in the multiplet become a Higgs triplet.
The nonlinear transformation of the Goldstone bosons under collective global
symmetries naturally ensures the absence of the SM Higgs mass term of the
form $m^2|h|^2$ at a TeV energy scale.
The Higgs mass squared parameter arises from the Coleman-Weinberg
potential in the gauge sector as well as in the fermion sector
at the electroweak scale\cite{Coleman:1973jx}.

In the LH model the hierarchy problem of the SM is naturally solved
in a similar way as in supersymmetric extensions of the SM:
The one-loop divergences from the SM particles of spins $J=1,1/2,0$
on the Higgs mass parameter are cancelled by those 
from new massive particles of the same spins $J=1,1/2,0$,
respectively\cite{Schmaltz:2002wx}.
Among the new heavy particles the fermion with spin $J=1/2$
is a vector-like heavy quark that is analyzed in this paper.
The existence of the heavy quark introduces new effects in the weak currents.
The CKM matrix is extended to $4\times3$ and flavor changing neutral 
currents occur at tree level.

This article is organized as follows.
In section 2 we review the LH model. We describe the gauge
sector as well as the fermion sector in detail. 
In section 3 we study the charged currents in the LH model 
and introduce the extended $4\times 3$ CKM matrix.
In section 4 we discuss the flavor changing nuetral currents
in the LH model and the neutral mixing angles.
We present a few of  measurements which determine the mixing
angles. In section 5 we draw a conclusion. We present the detailed
derivation of the neutral mixing anlges from the up-type
quark mass matrix and the Yukawa couplings in the Appendix.

\section{The Littlest Higgs Model}

The Littlest Higgs (LH) model begins with $SU(5)$ global symmetry, with a locally 
gauged subgroup $[SU(2)\times U(1)]^2$ \cite{Arkani-Hamed:2002qy}.
The $SU(5)$ global symmetry is spontaneously broken down to its subgroup
$SO(5)$ at the scale $f\sim$ 1 TeV. The vacuum expectation value
associated with the spontaneous symmetry breaking is
proportional to the $5\times 5$ symmetrical matrix                                                                                
\begin{equation}
\Sigma_0=\left(\begin{array}{ccc}
 & & {\mathbf 1}_{2\times2} \\
 & 1 &  \\
{\mathbf 1}_{2\times2} & & \end{array}\right).
\end{equation}
The global symmetry breaking results in fourteen massless Goldstone
bosons. Among them four massless Goldstone bosons are eaten
by the gauge bosons so that the gauge group  $[SU(2)\times U(1)]^2$
is  broken down to its diagonal subgroup $SU(2)\times U(1)$.
The remaining ten Goldstone bosons can be parameterized
by a non-linear field $\Sigma$ as follows:                                                                              
\begin{equation}
\Sigma=e^{i\Pi/f}\Sigma_0 e^{i\Pi^T/f}
=e^{2i\Pi/f}\Sigma_0
\end{equation}                                                                             
\begin{equation}
\Pi=\left( \begin{array}{ccc}
 & \frac{h^\dagger}{\sqrt{2}} & \phi^\dagger \\
\frac{h}{\sqrt{2}} & & \frac{h^\ast}{\sqrt{2}} \\
\phi & \frac{h^T}{\sqrt{2}} & \end{array} \right)
\end{equation}
%%%
Here $\Pi$ is the Goldstone bosons which fluctuate about
this background in the broken directions.
The field contents are grouped as a complex doublet $h$ and
a complex triplet $\phi$ with hypercharge $Y_h=1/2$ and $Y_\phi=1$
under the unbroken  gauge group $SU(2)_L\times U(1)_Y$:                                                                          
\begin{equation}
h = \left(\begin{array}{cc} h^+ &  h^0 \end{array} \right)\,,
\quad
\phi=\left( \begin{array}{cc} \phi^{++} & \frac{\phi^+}{\sqrt{2}}
\\ \frac{\phi^+}{\sqrt{2}} & \phi^0 \end{array} \right).
\end{equation}

The complex doublet becomes the SM Higgs doublet 
while the complex triplet is an addition to the SM particle contents.
The triplet acquires a TeV scale mass at one-loop from the Coleman-Weinberg
potential in the gauge sector as well as in the fermion sector.
The SM Higgs doublet acquires the mass squared parameter at two-loop
as well as at one-loop from the Coleman-Weinberg potential.
On the other hand, the SM Higgs quartic self-coupling arises after
integrating over the massive triplet.
The one-loop quadratic divergence on the SM Higgs mass parameter
from the SM Higgs self coupling is naturally cancelled 
by that from the triplet coupling to the SM Higgs doublet.
                                                                                                                               
\subsection{Gauge sector}
\label{Sec:Gauge}
The LH model has a broken $SU(2)\times U(1)$ symmetry 
as well as the unbroken Standard Model $SU(2)_L\times U(1)_Y$
symmetry at the scale $f$.
This is a key feature of the Little Higgs construction. 
Though the pseudo Goldstone multiplet vary from model to model 
all the Little Higgs models have extended gauge groups. 
The quadratic divergence of the SM gauge bosons in the Higgs mass
is cancelled by that of a heavy copy of the standard model gauge boson. 
On the other hand, In the sypersymmetric extension of the SM model 
the divergence of the Standard Model gauge bosons in the Higgs mass
is cancelled by the gaugino contribution.

The gauge group structure in the LH model is given as follows.
The generators of the $SU(2)$'s are embedded into $SU(5)$ as 
\begin{equation}
Q^a_1=\left( \begin{array}{cc} \sigma^a/2 & \quad \\ 
\quad  & {\mathbf 0}_{3\times 3} \end{array} \right),  \quad 
Q^a_2=\left( \begin{array}{cc} {\mathbf 0}_{3\times 3} & \quad \\ 
\quad  & \sigma^{a\ast}/2 \end{array} \right)
\end{equation}
%%%
while the generators of the $U(1)$'s are given by
\begin{equation}
Y_1=\left(\begin{array}{cc}
-\frac{3}{10}{\mathbf 1}_{2\times 2} & \quad \\
\quad & \frac{2}{10}{\mathbf 1}_{3\times 3}
\end{array}\right), \quad 
Y_2=\left(\begin{array}{cc}
-\frac{2}{10}{\mathbf 1}_{3\times 3} & \quad \\
\quad & \frac{3}{10}{\mathbf 1}_{2\times 2}
\end{array}\right).
\end{equation}
%%%
The generators of the electroweak symmetry $SU(2)_L\times U(1)_Y$ are 
expressed by $Q^a=\frac{1}{\sqrt{2}}(Q^a_1+Q^a_2)$ and $Y=Y_1+Y_2$.
The kinetic term for the pseudo Goldstone bosons can be written as
\begin{equation} {\cal L}_\Sigma=\frac{1}{2}\frac{f^2}{4} \mbox{Tr}\, 
|{\cal D}_\mu \Sigma|^2,
\end{equation}
where the covariant derivative is given by
\begin{equation}
{\cal D}_\mu \Sigma= \partial_\mu \Sigma +i\sum^2_{j=1}
[g_j(W_j\Sigma+\Sigma W^T_j)+g'_j(B_j \Sigma+\Sigma B^T_j)],
\end{equation}
with the gauge bosons are defined by
\begin{equation}
W_j=\sum^3_{a=1} W^a_{\mu j} Q^a_j,\quad B_j=B_{\mu j}Y_j.
\end{equation}

The spontaneous gauge symmetry breaking gives rise to mass terms
of order $f$ for the broken gauge bosons
\begin{eqnarray}
{\cal L}_{\Sigma,mass} 
&=& \frac{1}{2}\frac{f^2}{4} [g_1^2 W^a_{1\mu}W^{a\mu}_1
+g^2_2 W^a_{2\mu}W^{a\mu}_2-2g_1g_2 W^a_{1\mu}W^{a\mu}_2 ] \nonumber \\
& &+\frac{1}{2}\frac{f^2}{4}\frac{1}{5}[g'^2_1 B_{1\mu}B^\mu_1+g'^2_2
B_{2\mu}B^\mu_2-2g'_1g'_2 B_{1\mu}B^{2\mu}]
\end{eqnarray}
%%%
The $W, W_H, B$ and $B_H$ fields in mass eigenstates are defined as 
\begin{eqnarray} 
W &=& \sin \theta W_1+\cos\theta W_2,\qquad 
W_H=-\cos\theta W_1+\sin\theta W_2 \nonumber \\
B&=&\sin\theta' B_1+\cos\theta' B_2, \qquad 
B_H=-\cos\theta' B_1+\sin\theta' B_2
\end{eqnarray}
where the mixing angles are given by
\begin{eqnarray}
\sin\theta &=& \frac{g_2}{\sqrt{g^2_1+g^2_2}}, \qquad
\cos\theta= \frac{g_1}{\sqrt{g^2_1+g^2_2}} \nonumber \\
\sin\theta' &=& \frac{g'_2}{\sqrt{g'^2_1+g'^2_2}},\qquad
\cos\theta'= \frac{g'_1}{\sqrt{g'^2_1+g'^2_2}}.
\end{eqnarray}
At the energy scale $f$, the SM guage bosons $W$ and $B$ are massless 
and the masses of the heavy gauge bosons $W_H$ and $B_H$ are then given by
\begin{equation}
M_{W_H}=\frac{f}{2}\sqrt{g^2_1+g^2_2},\qquad
M_{B_H} =\frac{f}{2\sqrt{5}}\sqrt{g'^2_1+g'^2_2}.
\end{equation}

\subsection{Fermion Sector}
\label{Sec:Fermion}
In the Standard Model the Yukawa couplings for all fermions 
besides the top quark are small, and 
it is not necessary to protect the Higgs mass from their one-loop quadratic divergences.
The top Yukawa coupling is the strongest one-loop quadratic 
divergence in the Standard Model, 
so its negative contribution to one-loop quadratic divergence dominates 
all the positive contributions from the gauge sector as well as from the Higgs sector
in the Standard Model.

In the LH model we incorporate dominance of the top quark one-loop divergence
in the Standard Model with the introduction of the Higgs triplet and 
new heavy gauge bosons in order to trigger the electroweak symmetry breaking.
The Higgs doublet is a Goldstone boson at the scale $f$ 
so that there is no mass term for the Higgs doublet at the tree level.
The Higgs potential arises from the Coleman-Weinberg potential.
Additional heavy gauge bosons and the Higgs triplet give the Higgs
doublet a logarithmically  enhanced positive mass squared. 
In order to cancel out these positive contributions and get a negative Higgs mass
squared we introduce an additional quark 
in vector-like representation of the Standard Model gauge group\cite{Nir:1999mg},
\begin{equation}
\tilde t (3,1)_{+2/3} \quad +\quad \tilde t^c (\bar 3,1)_{-2/3}.
\end{equation}
Note that it is a singlet under the $SU(2)_L$ gauge group
to evade the gauge anomalies.

The LH model contains Yukawa couplings between the fermions
and the scalars. A large top Yukawa coupling arises 
because the Weyl fermions $\tilde t,\tilde t^c$ mix mostly 
with the usual third-generation weak doublet
$q_3=(b_3,t_3)$ and weak singlet $u'^c_3$.
The interaction between the $\Sigma$ field and quarks
in the up sector can be taken as 
\begin{equation}
{\cal L}_{U,mass}=\frac{1}{2}\lambda^U_{ab} f \epsilon_{ijk}\Psi_{ai}
\Sigma_{jx}\Sigma_{ky}u'^c_b+\lambda_0 f \tilde t \tilde t^c + h.c.
\end{equation}
\begin{equation}
\Psi_{1i}=\left(\begin{array}{c} d_1 \\ u_1 \\ 0 \end{array} \right),\quad
\Psi_{2i}=\left(\begin{array}{c} s_2 \\ c_2 \\ 0 \end{array} \right),\quad
\Psi_{3i}=\left(\begin{array}{c} b_3 \\ t_3 \\ \tilde t \end{array} \right)
\end{equation}
where $\lambda^U_{ab}$ are the ordinary Yakawa couplings in the up sector
and $\lambda_0$ is a new Yukawa coupling.
This is a slight but natural generalization 
of the scalar couplings to the quark in the up sector. 
Note that mixing between different generations occurs. 
Expanding the $\Sigma$ field generates the Higgs interactions with quarks.
After the Higgs doublet  gets a vev $v$ the up-type quark mass matrix 
${\cal M}_U$ becomes
\begin{equation}
{\cal M}_{U} =\left( \begin{array}{cccc}
i\lambda_{11} v& i\lambda_{12} v & i\lambda_{13} v  & 0 \\
i\lambda_{21} v  & i\lambda_{22} v & i\lambda_{23} v & 0 \\
i\lambda_{31} v & i\lambda_{32} v  & i\lambda_{33} v  & 0 \\
0 & 0 & \lambda_{33} f & \lambda_0 f
\end{array} \right) .
\end{equation}
Here we ignore the triplet vev $v'$ because its value is much smaller 
than the doublet vev, $v$\cite{Han:2003wu,Chen:2003fm}.  
Note that the element (3,4) of the matrix ${\cal M}_U$  is zero 
because there is no mixing between $t_3$ and $\tilde t^c$ in the lagrangian.  
The weak eigenstate $U_L$ and mass eigenstate $U^m_L$ are related by
$U^m_L=T^U_L U_L$ with $U^m_{L}=(u_L, c_L, t_L, T_L)^T$
and $U_{L}=(u_1, c_2, t_3, \tilde t )^T$.
The Higgs interactions with the down-type quarks are generated 
by a similar Lagrangian, again without the extra quarks.
  
\section{Charged Currents}
\label{Sec:CC}

The presence of the extra $\tilde t$ quark modifies the electroweak currents.
Now that the number of up-type quarks is four the matrix relating the 
quark mass eigenstates with the weak eigenstates  
becomes a $4\times3$ matrix in the LH model.
The Standard Model quark doublets couple only to the gauge group $SU(2)_1$. 
The charged currents in the LH model are given by\cite{Han:2003wu}
\begin{eqnarray}
{\cal L}_{CC}
&=&\frac{g_1}{\sqrt{2}}W_1^{\mu+} \bar{U}_L\gamma_{\mu}D_L +h.c.\nonumber \\
&=&\frac{g}{\sqrt{2}}W^{\mu+}  V_{ij}\bar{U}^{m}_{iL}\gamma_{\mu} D^m_{jL} 
-\frac{g}{\sqrt{2}\tan \theta} W_H^{\mu+} V_{ij}\bar{U}^m_{iL}\gamma_{\mu}D^m_{jL}
\nonumber \\ 
& &+\,{\cal O}\bigg(\frac{v^2}{f^2}\bigg)+h.c.
\end{eqnarray}
Here $V_{ij}$ is elements of the ``extended'' Cabibbo Kobayashi
Maskawa (CKM) matrix where the subindices $i(j)$ run over 1 to 4(3), 
the number of generations in the up(down) sector.  
$D_L$ and $D^m_L$ are left-handed quarks in the down sector in weak 
and mass eigenstates respectivley.
They are related by $U^m_{D} =T^D_L D_L$ with 
$D_{L}=(d_1, s_2, b_3)^T$ and $D^m_{L}=(d_L, s_L, b_L)^T$.
The electroweak coupling constant is defined as $g=g_1\sin\theta$.
In Eq.(18) we ignore small corrections arising 
from the Higgs vev.
We are not concerned with it in this case 
because the corrections do not change the extended CKM matrix.
Note that there are new charged currents associated with 
heavy gauge bosons $W_H$, and whose coupling constant is $g/\tan\theta$.
A heavy copy of the SM gauge bosons causes interference effects, 
which are suppressed by $M^2_W/M^2_{W_H}$
compared with processes mediated only by the weak gauge boson $W$.
For simplicity, we ignore the charged currents associated with 
the heavy gauge bosons $W_H$. 

Now we find the expression for the extended CKM matrix 
$V=(T^U_L)^\dagger T^D_L$.
It is convenient to work in a basis where the down-type quark mass 
eigenstates are identified with the weak eigenstates by setting 
$T^D_L={\bf 1}$ (unit matrix). 
The extended CKM matrix is then expressed by the elements of 
the up-type quark transformation matrix $T^U_L$ only:
\begin{equation}
V^{LH}_{ij} =(T^U_L)_{ij} \quad \mbox{for} \quad i=1,2,3,4\,; 
\quad j=1,2,3
\end{equation}
%%%
From now on we express the up-type quark transformation matrix 
$T^U_L$ as follows:
%%%
\begin{equation}
T^U_L=\left( \begin{array}{cccc} V_{ud} & V_{us} & V_{ub} & \Theta_u \\
V_{cd} & V_{cs} & V_{cb} & \Theta_c \\
V_{td} & V_{ts} & V_{tb} & \Theta_t \\
V_{Td} & V_{Ts} & V_{Tb} & \Theta_T 
\end{array} \right)
\end{equation}

Compared to the CKM matrix in the Standard Model 
the extended CKM matrix has the fourth row elements 
$V_{Td}$, $V_{Ts}$ and $V_{Tb}$.
These parameters can be measured by the decays of mesons 
composed of the $T$ quark and the down-type quarks. 
The inclusive decay rate $T\to q\ell \bar \nu$ is given by
\begin{equation}
\Gamma(T\to X_q \ell \bar \nu) \approx
\frac{G_F^2|V_{Tq}|^2}{192\pi^3}m^5_T
\approx 23\times|V_{Tq}|^2\bigg[\frac{m_T}{1\,\mbox{TeV}}\bigg]^5
\mbox{ GeV},
\end{equation}
where $q$ is down-type and $m_T$ is the $T$ quark pole mass.
For the b quark the value of $V_{Tb}$ is estimated 
in Ref.\cite{Han:2003wu}.
\begin{equation}
|V_{Tb}|\sim \frac{|\lambda_{33}|^2}{|\lambda_{33}|^2+|\lambda_0|^2}
\frac{v}{f}
\end{equation}
For $\lambda_{33}\sim \lambda_0\sim 1$ and $f\sim 1$ TeV,
the decay rate $T\to b\ell \bar \nu$ is given by 
\begin{equation}
\Gamma(T\to b \ell \bar \nu) \sim
0.023\times\bigg[\frac{m_T}{1\,\mbox{TeV}}\bigg]^5
\mbox{ GeV}.
\end{equation}
The dominant partial decay widths of the T quark are given in Ref.\cite{Han:2003wu}
\begin{equation}
\Gamma(T\to t h) =\Gamma(T\to tZ)=\frac{1}{2}\Gamma(T\to bW^+)
=\frac{1}{32\pi}\frac{|\lambda_{33}|^4}{|\lambda_{33}|^2+|\lambda_{0}|^2}m_T.
\end{equation}
Other decay modes are effectively suppressed by $v^2/f^2$.
The branching ratio of the inclusive decay $T\to b\ell \nu$ is given by
\begin{equation}
\mbox{Br}(T\to b\ell \nu) \sim 1.2\times10^{-3}\times
\bigg[\frac{m_T}{1\,\mbox{TeV}}\bigg]^4.
\end{equation}

\section{Neutral Currents}
\label{Sec:NC}
In the Standard Model the neutral-current interactions do not change 
the flavor at the tree level.
The smallness of the flavor-changing neutral currents(FCNC) 
transitons in the Standard Model is due to quantum loop effect
and the GIM cancellation mechanism. 
In contrast, the FCNC in the LH model occurs at the tree level
by the lagrangian\cite{Han:2003wu}
\begin{equation}
{\cal L}_{NC} =\frac{g}{\cos \theta_W}Z_{\mu}\,(J^{\mu}_{W^3}
-\sin^2 \theta_W J^{\mu}_{EM})
+ \frac{g}{\tan\theta} Z_{H\mu} J^{\mu}_{W^3} 
+ {\cal O}\bigg(\frac{v^2}{f^2}\bigg),
\end{equation}
where $J^\mu_{EM}$ is the electric current which is the same 
as in the Standard Model, 
and the weak neutral current $J^\mu_{W^3}$  is given by
\begin{eqnarray} J^{\mu}_{W^3} 
& = & \frac{1}{2}{\bar U}_L\gamma^\mu U_L
-\frac{1}{2} \bar{D}_L\gamma^\mu D_L \nonumber \\
& = & \frac{1}{2}{\bar U}^m_L\gamma^\mu \,\Omega\, U^m_L
-\frac{1}{2} \bar{D}^m_L\gamma^\mu D^m_L. 
\end{eqnarray} 
Note that the neutral currents in the down sector remains the same 
as those in the Standard Model while those in the up sector 
have additional currents associated with the $T$ quark.
The up-type quark transformation matrix generates 
a $4\times4$ neutral currents mixing matrix $\Omega$ in the up sector. 
The elements of $\Omega$ is then expressed only by the fourth column 
elements of $T^U_L$:
\begin{eqnarray}
\Omega & = & T^U_L\, \mbox{diag}\, (1,1,1,0) \,T^{U\dagger}_L  \nonumber \\
& & \nonumber \\
& =& \left( \begin{array}{cccc} 
1-|\Theta_u|^2 & -\Theta_u\Theta^*_c & -\Theta_u\Theta^*_t  & -\Theta_u\Theta^*_T \\
-\Theta_c\Theta^*_u & 1-|\Theta_c|^2 & -\Theta_c\Theta^*_t & -\Theta_c\Theta^*_T \\
-\Theta_t\Theta^*_u & -\Theta_t\Theta^*_c & 1-|\Theta_t|^2 & -\Theta_t\Theta^*_T \\
-\Theta_T\Theta^*_u & -\Theta_T\Theta^*_c & -\Theta_T\Theta^*_t  & 1-|\Theta_T|^2 
\end{array} \right)
\end{eqnarray}
Note that the diagonal elements are less than 1. 
Furthermore, it holds that $1-|\Theta_i|^2=|V_{id}|^2+|V_{is}|^2+|V_{ib}|^2$ for $i=u,c,t,T$ 
due to unitarity of the matrix $T^U_L$.  
From the Particle Data Book we have the upper bound on $|\Theta_i|$: 
$|\Theta_u| < 0.091,\, | \Theta_c|<0.147, \, |\Theta_t|<0.997$. 

Unitarity of the up-type transformation matrix $T^U_L$ also 
requires that the off-diagonal elements do not vanish. 
The Standard Model unitarity triangle should be replaced 
by a unitary quadrangle in the LH model.
For eaxmple, unitarity applied to the first and the second column 
yields
\begin{equation}
V_{ud}V_{cd}^*+V_{us}V_{cs}^*+V_{ub}V_{cb}^*+\Theta_u\Theta^*_c=0. 
\end{equation}
%%%
Then the last term on the left side becomes $\Omega_{uc}$. 
\begin{equation} 
\Omega_{uc}=-\Theta_u\Theta^*_c
= V_{ud}V_{cd}^*+V_{us}V_{cs}^*+V_{ub}V_{cb}^* \neq 0
\end{equation}

There is a way of estimating the absolute value of
$\Theta_{i},\,\,(i=u,c,t,T)$ from up-type quark 
mass matrix ${\cal M}_U$.
The details of the calculation are given in Appendix. 
Note that the values of $|\Theta_i|$ are related to 
the third column of the up-type quark transformation
 matrix $T^U_L$.
\begin{eqnarray}
|\Theta_u|&\sim & |V_{ub}|\frac{v}{f} \sim 0.001 \\
|\Theta_c|&\sim &|V_{cb}|\frac{v}{f} \sim 0.01\\
|\Theta_t|&\sim &|V_{tb}|\frac{v}{f} \sim 0.25\\
|\Theta_T|& \sim & 0.93
\end{eqnarray}
The last relation comes from the unitarity of the matrix 
$T^U_L$: $|\Theta_u|^2+|\Theta_c|^2+|\Theta_t|^2
+|\Theta_T|^2 =1$. The numerical values are estimated
for $f\sim 1$ TeV. The off-diagonal elements of the matrix 
$\Omega$ are then given by
\begin{eqnarray}
|\Omega_{uc}|&=& |\Theta_u||\Theta_c|\sim 10^{-5} \label{eq:uc}\\
|\Omega_{tu}|&=& |\Theta_t||\Theta_u|\sim 2.5\times 10^{-4}\label{eq:tu}\\
|\Omega_{tc}|&=& |\Theta_t||\Theta_c|\sim 2.5\times 10^{-3}. \label{eq:tc}
\end{eqnarray}

An interesting feature is that the up-type quark anomalous couplings
allow the Z-mediated FCNC at tree level only in the up sector.
This may modify the SM predictions for rare $D$ meson decays,
$D^0-\bar D^0$ mixing or $c\rightarrow u$ penguin processes.
This can also provide the possibility of rare top quark decays 
and same sign top pair production  
at the LHC as a direct probe of the Z-mediated FCNC.
In what follows, we consider the processes in the framework of the LH
model.

\subsection{Rare $D$ meson decays}
\label{Sec:RareD}
Hadronic $D$ meson decays are completely dominated by nonperturbative
physics and in general do not constitute a suitable test of the short 
distance structure of the Standard Model. 
But leptonic modes such as $D^0\rightarrow l^+ l^-$ can be used 
to constrain the size of the Z-mediated FCNC couplings in any extension
of the Standard Model. In the LH model the SM gauge boson $Z$
coupling $\Omega_{uc}Z \bar u_L c_L$ can give rise to the decays
$D^0\to e^+e^-,\,\mu^+\mu^-$ at tree level. 
The contribution of the $Z$-mediated tree level diagram to the branching ratio
of $D^0\to \mu^+\mu^-$, normalized to that of the $W$-mediated 
$D^+\to \mu^+\nu$, is given by 
\begin{equation} 
\frac{\mbox{Br}(D^0\to \mu^+\mu^-)_Z}{\mbox{Br}(D^+\to\mu^+\nu)} 
=2\bigg[\frac{\tau(D^0)}{\tau(D^+)}\bigg] [(1/2-\sin^2\theta_W)^2+\sin^4\theta_W] 
\frac{|\Omega_{uc}|^2}{|V_{cd}|^2}.
\end{equation}
Using eq.~(\ref{eq:uc}) we estimate the ratio:
\begin{equation}
\frac{\mbox{Br}(D^0\to \mu^+\mu^-)_Z}{\mbox{Br}(D^+\to \mu^+\nu)} 
\sim 2\times10^{-10}
\end{equation} 
Using the experimental bound 
Br$(D^+\to \mu^+\nu)=(3.5\pm1.4)\times10^{-4}$\cite{Besson:2004bs}
the branching ratio of $D^0\to\mu^+\mu^-$ is given by
\begin{equation}
\mbox{Br}(D^0\to\mu^+\mu^-)\sim 7\times 10^{-14}.
\end{equation}
The estimated SM branching ratio is $\sim 10^{-19}$\cite{Gorn:1978sb}, and 
the present experimental bound is $1.3\times 10^{-6}$\cite{Egede:2004ru},
still undetectable by foreseeable experments.

\subsection{$D^0-\bar D^0$ mixing }
\label{Sec:DDmix}
In this section we discuss how the LH model affects $D^0-\bar D^0$
mixing. The recent experiments confirm that $D^0-\bar D^0$ mixing
proceeds very slowly \cite{Aubert:2003ae}.
The Stanard Model explains that short distance contributions to
$D^0-\bar D^0$ mixing occurs via box diagrams, and the $s$-quark
contribution in the box diagrams is dominant.
The $b$-quark contribution is much smaller due to the small CKM elements 
$|V_{ub}V^\ast_{cb}|^2/|V_{us}V^\ast_{cs}|^2={\cal O}(10^{-6})$,
and due to the relative smallness of $m_b$ \cite{Georgi:1992as,
Burdman:2003rs}. 
That is, the Glashow-Ilipoulos-Maiani (GIM) cancellation is very effective
in the $D^0-\bar D^0$ system compared with that 
in the $B^0-\bar B^0$ system.
In contrast, long distance contributions to $D^0-\bar D^0$  mixing
are inherently nonperturbative, and cannot be calculated from 
the first principles.
Interestingly, mixing from the long distance effects can dominate over
the short distance ones, and even reach the current experimental
limit \cite{Falk:2001hx,Falk:2004wg}.

\begin{figure}[htb]
\centerline{
\hbox{\hspace{1cm}
\includegraphics[width=5.5truecm]{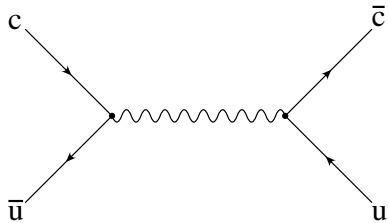}}}
{\caption[1]{\label{fig:DDmixing}
Tree-level FCNC $D^0-\bar D^0$ mixing diagram contributes only to the $x$
parameter.}}
\end{figure}

We now review the current experimental data for charm
mixing. Two physical parameters that characterize the mixing are
\begin{equation}
x=\frac{\Delta m}{\Gamma}, \qquad 
y=\frac{\Delta \Gamma}{2\Gamma}
\end{equation}
where $\Delta m$ and $\Delta \Gamma$ are mass and width differences of
the two neutral $D$ meson mass eigenstates and $\Gamma$ is their
averaged width. 
The present upper bounds on $x$ and $y$ are at a few times $10^{-2}$ level
\cite{Aubert:2003ae}.
In Ref. \cite{Falk:2001hx}, it is asserted that the values of $|x|$ and $|y|$ are 
between $10^{-3}$ and $10^{-2}$ within the Standard Model.
To check theoretical prediction more
precisely, one need to improve sensitivity at least at 0.1\% level in the
future generation of charm collider experiments\footnote{
Due to uncertainty from the Standard Model prediction
for charm mixing, the only robust potential signal of new physics
in charm system at this stage comes from  large CP violation.}.
In some cases,
even the current experimental limit is enough to severely constrain
new physics beyond the Standard Model.

In the LH model, the anomalous coupling $\Omega_{uc}$ produces 
$D^0-\bar D^0$ mixing at tree level by the $Z$ boson exchange,
as depicted in Fig. \ref{fig:DDmixing}, and
results in the short distance $\Delta C=2$ transition contributing only to $x$.
In what follows, we compute only the LH model contribution.
The mass difference due to $Z$ boson $\Delta m_D^{Z}$ is given by

\begin{equation}
(\Delta m_D)^Z=\frac{\sqrt{2}}{3} G_F m_D f_D^2 B_D\eta_D 
|\Omega_{uc}|^2\label{eq:mdiff}
\end{equation}
where $m_D$ is mass of the $D^0$ meson,
$f_D$ is decay constant of the $D^0$ meson,
a QCD correction $\eta_D$ is $\sim0.8$,  
and $B_D =1$ in vaccum saturation approximation.
These give the expression
\begin{equation}
(\Delta m_D)^Z\approx 2\times10^{-7}\!\times |\Omega_{uc}|^2 
\mbox{ GeV}.
\end{equation}
For $|\Omega_{uc}| \sim 10^{-5}$, the 
mass difference is estimated to be $\sim 10^{-17}$ GeV, and
this corresponds to $x\sim 10^{-5}$. 
One notes that the Standard Model prediction for charm mixing is much larger
than the LH model prediction by at least  a factor of 100.
As for $y$, one expects that the $Z$ mediated process contributes
only to $\Delta C=2$ and the LH model does not give any
significant contribution to $y$.

\subsection{$t\to cZ$ decay}
\label{Sec:tZc}
The LHC will produce millions of top quarks at the detectors
and most top quarks decay to $bW^+$ by weak interactions.
There can also be rare top quark decays like $t\to cZ$ with a real $Z$ boson. 
In the SM it occurs at one-loop level, and its branching ratio
is predicted to be $\sim10^{-12}$\cite{Eilam:1990zc}.
In contrast, in the LH model $t\to cZ$ decay occurs at tree level and 
the decay width is given by 
\begin{eqnarray}
\Gamma(t\to cZ)& =&|\Omega_{tc}|^2\frac{g^2}{128\pi\cos^2\theta_W} \frac{m^3_t}{m^2_Z} 
\bigg(1-\frac{m^2_Z}{m^2_t}\bigg)^2\bigg(1+2\frac{m^2_Z}{m^2_t}\bigg)\nonumber \\
&\approx &1.0\times|\Omega_{tc}|^2\,\mbox{GeV}.
\end{eqnarray} 
Using eq.~(\ref{eq:tc}) the decay width is estimated by
\begin{equation}
\Gamma(t\to cZ)\sim 6\times 10^{-6}\mbox{ GeV}.
\end{equation}
Assuming that the total width of the top quark is dominated by $t\to bW^+$, 
the branching ratio of $t\to cZ$ decay in the LH model is given by
\begin{equation}
\mbox{Br}(t\to cZ)\approx
\frac{\Gamma(t\to cZ)}{\Gamma(t\to bW^+)}
\approx\frac{1}{2\cos^2\theta_W}
\frac{|\Omega_{tc}|^2}{|V_{tb}|^2}\frac{m^2_W}{m^2_Z}
\approx 3\times 10^{-6}.
\end{equation}
By ignoring the up and charm quark masses and replacing the anomalous 
couplings, one can compute the branching ratio of $t\to uZ$ decay. 
\begin{equation}
\mbox{Br}(t\to uZ)\sim 5\times10^{-11}.
\end{equation}
The sensitivity of ATLAS experiment to the FCNC decay $t\to Zq$ (with $q=u,c$)
has been analyzed \cite{Chikovani:2000wi} by searching for a signal in the channel
$t\bar t\to (Wb)(Zq)$, with the boson being reconstructed
via the leptonic decay $Z\to \ell\ell$. 
The $t\to Z(u,c)$ signal in the ATLAS 
should be very clean but, due to the low signal event rare
and large backgrounds, only $\sim 3\times 100$fb$^{-1}$ of 
integrated luminosity would allow one to probe Br($t\to Z(u,c))$ as
low as $10^{-4}$ \cite{Beneke:2000hk}. Therefore,
the LH model prediction for decay $t\to Zq$ can not be tested in the
ALTAS and CMS detectors.

\subsection{Like-sign tt ($\bar t\bar t$) pair production}
\label{Sec:ttpair}
\begin{figure}[htb]
\centerline{
\hbox{\hspace{1cm}
\includegraphics[width=5truecm]{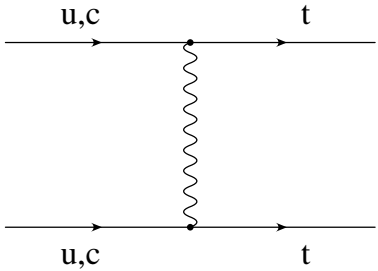}}}
{\caption[1]{\label{fig:ttproduct}
t-channel for same sign $tt$ production process 
(There is also u-channel diagram.)}}
\end{figure}

The LHC will begin to outline the physics at TeV energy scale. 
With an integrated luminosity of about 100 fb$^{-1}$ the LHC is expected 
to produce several tens of millions of $t\bar t$ pairs for each of the detectors, 
ATLAS and CMS per year\cite{Beneke:2000hk,Abe:1997fz,Cabrera:2003dp}.
Such a high rate of production will allow the LHC to search for new physics
associated with the top (anti-top) quark. For example, one take into account
a process in which two incoming up-type quarks exchange a neutral gauge
bosons like gluon, $\gamma, Z$
and then change to a pair of (same sign) top quarks\cite{Gouz:1998rk},
as depicted in Fig. \ref{fig:ttproduct}.

In this section, we consider the production cross section for $pp\to 
tt$ or $\bar t \bar t$ processes in the framework of the LH model.
The dimension 4 operators with the anomalous couplings 
$\Omega_{tq}\,\,(q=u,c)$ produce a signal of same sign top quark
pairs at tree level by the exchange of $Z$.

\begin{equation}
\frac{g}{2\cos \theta_W} Z_{\mu}
(\Omega_{tu}\, \bar t_L \gamma^\mu u_L
 +\Omega_{tc}\,\bar t_L\gamma^\mu c_L) + h.c.   
\end{equation}
To compute the production cross section for $pp\to tt$ process,
one takes into account $uu, uc, cc\to tt$
processes at the parton level. Due to the larger parton luminosity
from the $u$ quark one expect that the process $uu\to tt$ dominates over the
process $cc\to tt$. However, one note that the top-charm anomalous coupling
is larger than the top-up anomalous coupling by a factor of 10 in the
LH model. Thus we consider the $c$ quark as well as the $u$ quark.
Then the differential cross section for $pp\to tt$ process
is given in parton variables as follows
\begin{eqnarray}
\frac{d^3\sigma}{dx_1dx_2d\hat t}(pp\to tt+X)
&=&f_u(x_1)f_u(x_2) \frac{d\sigma}{d\hat t}(uu\to tt) \nonumber \\
&&\!\!\!+\,2f_u(x_1)f_c(x_2) \frac{d\sigma}{d\hat t}(uc\to tt) \nonumber \\
&&\!\!\!+\,f_c(x_1)f_c(x_2) \frac{d\sigma}{d\hat t}(cc\to tt).
\end{eqnarray}
where $x_i$ is the longitudinal fraction of the proton's momentum,
$\hat t$ is the conventional Mandelstam variable,
and $f_q(x)$ is the parton distribution function (pdf) of $q$ quark
in the proton.
There are the equivalent $\bar u \bar u, \bar u \bar c, 
\bar c \bar c \to \bar t \bar t$ like-sign anti-top processes.
The differential cross section for these processes
is given by
\begin{eqnarray}
\frac{d^3\sigma}{dx_1dx_2d\hat t}(pp\to \bar t\bar t+X)
&=&f_{\bar u}(x_1)f_{\bar u}(x_2) \frac{d\sigma}{d\hat t}
(\bar u\bar u\to \bar t\bar t) \nonumber \\
&&\!\!\!+\,2f_{\bar u}(x_1)f_{\bar c}(x_2) \frac{d\sigma}{d\hat t}
(\bar u\bar c\to \bar t\bar t) \nonumber \\
&&\!\!\!+\,f_{\bar c}(x_1)f_{\bar c}(x_2) \frac{d\sigma}{d\hat t}
(\bar c\bar c\to \bar t\bar t).
\end{eqnarray}

Here we do not directly compute the production cross sections.
Instead, we quote the result of the computation in Ref.\cite{Larios:2003jq}
where the anomalous coupling $a^L_{tq}$ are defined as
$a^L_{tq}\equiv\frac{g}{2\cos\theta_W} \Omega_{tq}$, and 
the plot in Fig.2 shows the LHC production cross section in parton levels.
For $|\Omega_{tu}|\sim 2.5\times10^{-4}$ and 
$|\Omega_{tc}|\sim 2.5\times10^{-3}$, the production cross section
for $pp\to tt$ process is less than 1 event for a 100 fb$^{-1}\,$data sample.
This rate is lower than the sensitivity of the LHC to the same
sign $tt(\bar t\bar t)$ pair production, and it is due to small top
quark anomalous couplings in the LH model. 

Note that converting a cross section measurement into a measurement of
the coupling constant requires knowledge of the pdfs in the proton.
The pdfs of the $c$ and $\bar c$ quarks in the proton coincide 
while those of $u$ and $\bar u$ quarks in the proton do not.
It causes the difference of production cross sections 
for the $pp\to tt$ process and for the $pp\to \bar t \bar t$ process.
These lead to direct measurement of the coupling constants 
$|\Omega_{tu}|$ and $|\Omega_{tc}|$. The numerical computation of the coupling
constants is found in Ref.\cite{Gouz:1998rk}. 
%Note that the cross section for the $cc\to tt$ process is suppressed by the pdf 
%of $c$ quark in the proton.

\section{Conclusion}

The Littlest Higgs model attributes the lightness of  the Higgs
to being a pseudo Goldstone boson,
and provides a simple mechanism for
electroweak symmetry breaking by introducing a heavy vector-like quark
to trigger a negative mass squared parameter for the Higgs doublet.
The presence of the new vector-like quark, as a consequence, modifies
the Standard Model predictions for the flavor neutral changing currents
as well as for the weak charged currents.
The extended $4\times3$ CKM matrix contains extra three mixing
parameters, while the neutral current mixing matrix consists
of four new complex parameters. These parameters induce a couple of 
rare processes for the up-type quarks. We have showed that 
the LH model predictions for the mixing angles are too low
compared to the current experimental bounds, and
the testability of the model requires more stringent 
measurements of the mixing angles at the ATLAS and CMS detectors
of the LHC in the future.
We conclude that the quark sector of the LH medel is not a 
suitable test of the signals of the LH medel. 
There are more complex little Higgs models which demand a heavy
vector-like quark just as the LH model does. 
As long as the models have only a  vector-like quark 
they might show the similar characteristic signatures in the
flavor neutral changing currents as well as in the charged currents.
To distinguish other Little Higgs models from the Littlest
Higgs Model, one should search for new signals in the gauge sector and
in the Higgs sector as well.

\section{Acknowledgements}

I would like to thank Ann Nelson for stimulating conversations. 
J.~L. was partially supported by the DOE under Contract DE-FG02-96ER40956, 
and the RRF research Grant from the University of Washington.

\bigskip

\appendix

\section{Elements of the unitary matrix $T^U_L$}
\label{sec:A}

The matrix $T^U_L$ can be computed by diagonalizing the Hermitian 
matrix ${\cal M}_U {\cal M}^\dagger_U$: 

\begin{equation}
{\cal M}^2_{diag}=
T^U_L \,{\cal M}_U {\cal M}^\dagger_U \,T^{U\dagger}_L
\end{equation}
where ${\cal M}_{diag}$ is given by

\begin{equation}
{\cal M}^2_{diag}=
\left(\begin{array}{cccc} 
m_u^2 & 0 & 0 & 0 \\
0 & m_c^2 & 0 & 0 \\
0 & 0 & m^2_t & 0 \\
0 & 0 & 0 & m^2_T
\end{array} \right)
\end{equation}
with $m_q$ is the $q$ quark mass.
The up-type quark mass matrix ${\cal M}_{U}$ is given by Eq.(17).  
The mass matrix squared ${\cal M}_U {\cal M}^\dagger_U$ is then
expressed as follows:
\begin{equation}
{\cal M}^2_{U,diag}= T_{U,L}\,{\cal M}_U {\cal M}^\dagger_U\,T^\dagger_{U,L}
\end{equation}
\begin{equation}
{\cal M}_U{\cal M}_U^\dagger =f^2
\left( \begin{array}{cccc}
 &  &  &  -i\lambda_{13} \lambda_{33}^* \frac{v}{f}  \\
 & {\bf \Lambda \Lambda^\dagger} (\frac{v}{f})^2  &  & -i\lambda_{23}\lambda_{33}^* \frac{v}{f}  \\
  &  &  & -i|\lambda_{33}|^2 \frac{v}{f}  \\
 i\lambda_{13}^* \lambda_{33} \frac{v}{f} & i\lambda_{23}^*\lambda_{33} \frac{v}{f}
 & i|\lambda_{33}|^2 \frac{v}{f} & |\lambda_{33}|^2+|\lambda_0|^2
\end{array} \right)
\end{equation}
where $(\Lambda)_{ij}=\lambda_{ij}$.
The matrix equation is so complicated that the perturbation method 
is a good way to solve it.
Then the problem is to solve eigenvalue problem in quantum mechanics 
with four eigenstates. 

Set $\frac{v}{f} \le \frac{1}{4}$ by choosing $f \ge 1$ TeV.
\begin{equation}
|\lambda_0| \sim |\lambda_{33}|  \gg  |\lambda_{ij}|
\quad \mbox{where} \,\, (i,j) \ne (3,3)
\end{equation} 
Each state is a mass eigenstate for the up-type quark 
and the coefficient of the states is the elements of $T_{U,L}$.
The unperturbed Hamiltonian is then given by
\begin{equation} H_0=f^2
\left( \begin{array}{cccc}
0  & 0  &  0 &  0  \\
0  & 0  &  0 &  0  \\
0  &  0 & |\lambda_{33}|^2(\frac{v}{f})^2& -i|\lambda_{33}|^2 \frac{v}{f}  \\
0 & 0 & i|\lambda_{33}|^2 \frac{v}{f} & |\lambda_{33}|^2+|\lambda_0|^2
\end{array} \right)
\end{equation}
and other elements are the perturbed Hamiltonian $H^\prime$.
Then the coefficients are given to leading order by 
\begin{equation} 
\Theta_u \approx \frac{\langle u|H^\prime|T\rangle}{m_u^2-m^2_T},
\end{equation}
where $|u\rangle$ and $|T\rangle$ are unpertured quark mass eigenkets,
and $m_u$ and $m_T$ are masses of $u$ and $T$ quarks respectively.
The value of $|\Theta_u|$ is then estimated as 
\begin{equation}
|\Theta_u| \sim \frac{|\lambda_{13} \lambda_{33}^*|}{ |\lambda_{33}|^2+|\lambda_0|^2}
\frac{v}{f}
\end{equation}
The value of $|V_{ub}|$ can be estimated in the same way:
\begin{eqnarray}
V_{ub} &\approx& \frac{\langle u|H'|t\rangle}{m_u^2-m_t^2} \\
|V_{ub}|&\sim& 
\frac{|\lambda_{13} \lambda_{33}^*|}{ |\lambda_{33}|^2+|\lambda_0|^2}
\frac{v^2}{f^2}
\end{eqnarray}
Therefore, the value of $|\Theta_u|$ is estimated from the $|V_{ub}|$.
\begin{equation}
|\Theta_u|\sim |V_{ub}|\frac{v}{f}
\end{equation}
Likewise,
\begin{eqnarray}
|\Theta_c| &\sim & |V_{cb}| \frac{v}{f} \\
|\Theta_t| &\sim & |V_{tb}| \frac{v}{f}
\end{eqnarray}
Then the anomalous coupling for the up and charm quarks 
is estimated as
\begin{equation}
|\Omega_{uc}|= |\Theta_u||\Theta_c|
\sim|V_{ub}||V_{cb}|\frac{v^2}{f^2}.
\end{equation}

\end{document}